\documentclass[fleqn,twoside]{article}
\usepackage{espcrc2}
\usepackage{epsfig}

\usepackage{graphicx}
\usepackage[figuresright]{rotating}

\newcommand{\AmS}{{\protect\the\textfont2
  A\kern-.1667em\lower.5ex\hbox{M}\kern-.125emS}}

\title{The scaling dimension of low lying Dirac eigenmodes and 
of the topological charge density}

\author{MILC Collaboration: C. Aubin\address[WU]{Department of
	Physics, Washington University, St. Louis, MO 63130, USA}
        C. Bernard\addressmark,
	Steven Gottlieb\address{Department of Physics, Indiana
        Univerity, Bloomington, IN 47405, USA},
	E.B. Gregory\address[AZ]{Department of Physics, University of
        Arizona, Tucson, AZ 85721, USA},
	Urs M. Heller\address{American Physical Society, One Research
        Road, Box 9000, Ridge, NY 11961, USA},
	J.E. Hetrick\address{Physics Department, University of the
        Pacific, Stockton, CA 95211, USA}\thanks{Presented by
        J.E. Hetrick},
	J. Osborn\address{Physics Department, University of Utah, Salt
        Lake City, UT 84112, USA},
	R. Sugar\address{Department of Physics, University of
        California, Santa Barbara, CA 93106, USA},
	D. Toussaint\addressmark[AZ] 
	with: Ph. de Forcrand\address{Institute for Theoretical 
	Physics, ETH Z\"urich, CH-8093 Z\"urich, Switzerland}\address{CERN, 
	Theory Division, CH-1211 Geneva 23, Switzerland},
	and O. Jahn\address{Center for Theoretical Physics, MIT,
        Cambridge, MA 02139, USA}}
       
\begin{document}

\begin{abstract}
As a quantitative measure of localization, the inverse participation
ratio of low lying Dirac eigenmodes and topological charge density is
calculated on quenched lattices over a wide range of lattice spacings
and volumes. Since different topological objects (instantons,
vortices, monopoles, and artifacts) have different co-dimension,
scaling analysis provides information on the amount of each present
and their correlation with the localization of low lying eigenmodes.
\vspace{1pc}
\end{abstract}

\maketitle

\section{INTRODUCTION}

With modern computational power has come the ability to examine the low
lying eigenvectors of the Dirac operator and hence their spatial
correlation with instantons and other related objects thought to be
involved in chiral symmetry breaking and
confinement \cite{Eigcoor,IPR}. While these studies focused
primarily on the local relationship between instantons and low-lying
Dirac eigenmodes (LDEs), other models of confinement and chiral
symmetry breaking involving objects of lower co-dimension are popular,
based on monopoles, vortices, and hybrid
objects \cite{Engl}. Presumably these objects would have a rather
different effect on the LDEs than 4-dimensional instantons, due to
their different co-dimension.  Furthermore, a recent study
\cite{Horvath} has suggested a dense layered 3-dimensional structure
to the LDEs.

One difficulty is the quantitative characterization of localization of
the LDEs or related quantities such as the topological charge
density. 
In \cite{IPR} localization of the LDEs was studied using the
inverse participation ratio (IPR) which yields a number
characterizing the localization of an eigenmode.

By studying the scaling dimension of the IPR, we can find the
co-dimension of the structures which localize the LDEs, thus giving
some insight as to the possible confining objects and mechanism.

\section{INVERSE PARTICIPATION RATIO (IPR)}

The IPR of a normalized field $\rho_i(x)$ is defined as
\begin{equation}
I = N \sum_x \rho_i^2(x)
\end{equation}
where $N$ is the number of lattice sites $x$. Here we use $\rho_i(x) =
\psi_i^\dagger\psi_i(x)$ and $\psi_i(x)$ is the $i$-th, normalized 
($\sum_x \rho_i(x) = 1$), lowest eigenvector of the Dirac operator.

With this definition, $I$ characterizes the inverse ``fraction'' of
sites contributing significantly to the support of $\rho(x)$ (we
now drop the subscript $i$). A simple calculation shows that the IPR
takes the following values for these simple situations:
\begin{eqnarray*}
{\rm Unlocalized:~} \rho(x) = {\rm const.}~ && I = 1\\
\delta-{\rm function:~} \rho(x) = \delta(x_o)~ && I = N\\
{\rm ~~localized~on~fraction~}f{\rm ~of~sites:} && I = 1/f
\end{eqnarray*}

Suppose that the objects responsible for confinement, or indeed any
physics governing the lowest Dirac eigenmodes, localize the LDEs. As
the lattice spacing is reduced, the fraction of sites contributing to
the IPR scales as $a^d/a^4$. Thus the IPR indicates the co-dimension
of these objects: $d=4$ for instantons, $d=3$ for monopoles, and $d=2$
for vortices.  Gauge dislocations should contribute as $d=0$ objects,
however their density diverges as $a^{-4}$ so that they should give a
$\sim$constant contribution: $a^0/(a^4 a^{-4})$).

Since the IPR $\sim 1/f$, if we reduce the lattice spacing at fixed
physical volume, we have
\begin{equation}
a \rightarrow 0~~{\rm at~fixed~volume:~~} I \sim a^{4-d}
\end{equation}
On the other hand, increasing the volume at fixed lattice spacing
includes proportionately more of the confining objects, whatever
their dimension. Thus we expect the IPR to remain constant,
\begin{equation}
L \rightarrow \infty~~{\rm at~fixed~}a:~~ I \sim {\rm constant}
\end{equation}

\section{RESULTS}

We have explored these two regimes using quenched lattices generated
with the tadpole improved Symanzik gauge action, and the parameter
set shown in Table 1.  On each lattice we computed the lowest eight
eigenvectors of the Asqtad Dirac matrix. 
\vspace{-0.5cm}
\begin{table}[htb]
\caption{Lattices analyzed}
\label{table:1}
\begin{tabular}{@{}|cccccc|}
\hline
\hline
& $a$ & $L$ & vol & $\beta$ & no. configs. \\
\hline
\hline
& $a \rightarrow 0$: & & & & \\ 
\hline
& 0.20 fm & 12 & (2.4 fm)$^4$ & 7.56  & 100\\
& 0.15~~~~    & 16 &     .        & 7.847 & 97\\
& 0.12~~~~    & 20 &     .        & 8.109 & 93\\
& 0.095~~~~   & 24 & (2.3 fm)$^4$ & 8.456 & 118\\
& & & & & \\
\hline
& $L \rightarrow \infty$: & & & &\\
\hline
& 0.12 fm & 12 & (1.4 fm)$^4$  & 8.109 & 100\\
&   .     & 16 & (1.9 fm)$^4$  & .     & 100\\
&   .     & 20 & (2.4 fm)$^4$  & .     & 93\\
&   .     & 24 & (2.9 fm)$^4$  & .     & 100\\
\hline
\end{tabular}
\end{table}

\vspace{-0.7cm}
We see clear evidence for lower
dimensional scaling as $a\rightarrow 0$
with co-dimension between 2 and 3.
Figure 1 shows both the distribution of
the IPRs and the scaling of the averages.
\begin{figure}[htb]
\vspace{-0.5cm}
\epsfxsize=8cm
\epsfxsize=8cm
\epsfbox[60 80 410 300 ]{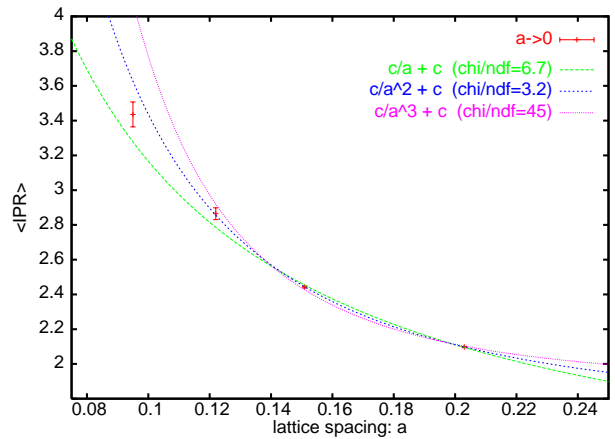}
\caption{Scaling of the average IPR as $a \rightarrow 0$}
\end{figure}
\vspace{-0.5cm}
The data points are fit to $c_1/a^n + c_2$, where $c_1$ and $c_2$ are
constants and $n = 1,2,3$.  The reduced chi squared
values\footnote{The error bars shown here are corrected from those
(much larger) shown at the conference.}  for the fits are 6.7, 3.2,
and 45 for $n = $1, 2, and 3, respectively.

In figure 2 we show the behaviour of the IPR as we increase the volume
at fixed lattice spacing; it is rather unaffected, as we expected above.
\begin{figure}[htb]
\vspace{-0.5cm}
\epsfxsize=8cm
\epsfxsize=8cm
\epsfbox[60 80 410 300 ]{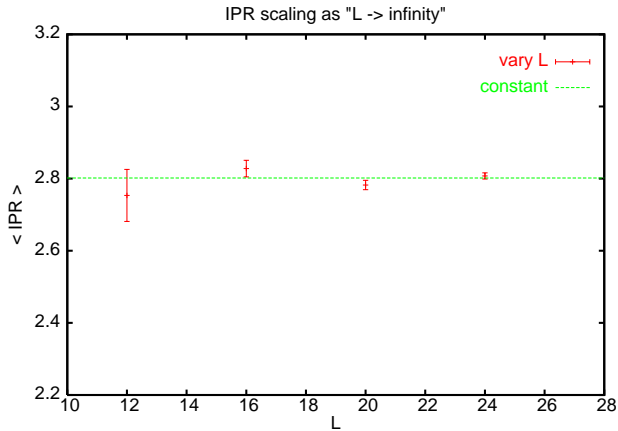}
\caption{Scaling of the average IPR as $L \rightarrow \infty$ with $a
= 0.12$ fm}
\end{figure}

\vspace{-1cm}
\section{TOPOLOGICAL CHARGE}

We can also investigate the localization of the topological charge
density by computing the IPR from $q(x)=F_{\mu\nu}\tilde F^{\mu\nu}$, 
where we have normalized $\sum_x |q(x)| = 1$. We have computed
$q(x)$ by successive HYP smearing sweeps \cite{Topo}
on the $a\rightarrow 0$ series, and show the results in Figure 3 (note
that only 5 HYP smearing steps were performed on the $a=0.12 {\rm
~fm~} L=20$ lattice set). While this plot does not show us new
information on the localization of the LDEs, it is nonetheless
instructive.

First, we see that all lattices without smoothing have an IPR =
$\pi/2$. This is the value expected if the field is a gaussian
fluctuation at each site, regardless of its width. We further see the
approach to a stable localization of topological charge versus HYP
smearing as the lattice spacing is decreased (at fixed volume). 
We note that $<$IPR$>$ is not
large, meaning that $q(x)$ is not strongly localized. Also, it
increases as $a \rightarrow 0$ as for the LDEs.

\begin{figure}[htb]
\vspace{-0.5cm}
\epsfxsize=8cm
\epsfbox[60 40 410 300 ]{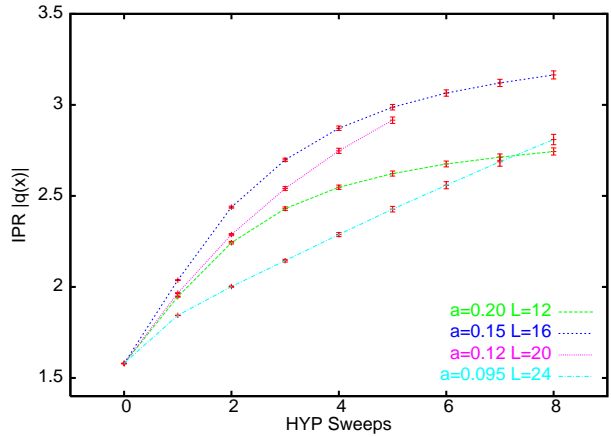}
\vspace{-1.5cm}
\caption{Avereage topological IPR vs. number of HYP smearing sweeps,
as $a\rightarrow 0$}
\end{figure}
\vspace{-0.3cm}
\section{CONCLUSIONS}
The main result of this study is the indication
of a localization of the low-lying Dirac eigenmodes on
surfaces of co-dimension between 2 and 3, qualitatively supporting the
center vortex or monopole pictures of confinement. 
Note, however, that the singularities of {\em thin} objects (vortices
or monopoles) are expected to be smoothed out by the QCD interactions
and become {\em thick}, with a size $\sim 1/\Lambda_{\rm QCD}$. Thick objects
fill a fixed fraction of space, not a divergent one.
The indication we have, via the divergence of $<$IPR$>$ as $a\rightarrow 0$, of
localization on singular manifolds, is remarkable, whatever these
manifolds are.




\begin{thebibliography}{9}
\bibitem{Eigcoor} Ph. de Forcrand et. al., Nucl. Phys. Proc. Suppl. {\bf
73} (1999) 578; T. DeGrand and A. Hasenfratz, Phys. Rev. {\bf D64} 
(2001) 034512; T.L. Ivanenko and J.W. Negele,
Nucl. Phys. Proc. Suppl. {\bf 63} (1998) 504
\bibitem{IPR} C. Gattringer et. al., Nucl. Phys. {\bf B617} (2001) 101;
T. Kovacs, Phys. Rev. {\bf D67} (2003) 094501
\bibitem{Engl} M. Engelhardt, these proceedings.
\bibitem{Horvath} I. Horvath et. al., Phys. Rev. {\bf D68} (2003) 114505;
H.B. Thacker, these proceedings.
\bibitem{Topo} T. DeGrand, A. Hasenfratz, T.G. Kovacs Nucl. Phys. {\bf
B505} (1997) 417
\end{thebibliography}
\end{document}